\documentstyle[preprint,revtex,eqsecnum]{aps}

\begin{document}
\draft
\preprint{Alberta-Thy-5-93}
\begin{title}
Dynamics and Collision of Massive Shells in Curved Backgrounds
\end{title}
\author{ Dar\'\i o N\'u\~nez$^{\dagger \ast}$,

H. P. de Oliveira$^{\dagger \ast \ast}$, Jose Salim$^{\dagger
\ddagger}$  }

\begin{instit}
$^{\dagger}$ Theoretical Physics Institute, University of Alberta,
Edmonton, \\
Alberta T6G 2J1, Canada. \\

$^{\ast}$ Instituto de Ciencias Nucleares, UNAM, \\
CU, A.P.

70-543, M\'exico, D. F. 04510,  M\'exico. \\

$^{\ast \ast}$Universidade do Estado do Rio de Janeiro, Instituto de
Fisica, \\ R. Sao Francisco Xavier, 524, Maracana, CEP 20550, Rio de
Janeiro. \\

$^{\ddagger}$ Centro Brazileiro de Pesquisas Fisicas. \\
Sigaud 150, Rio du Janeiro, Cep 22290,  Brazil.
\end{instit}

\begin{abstract}
{\it We analyse the dynamics of the collision of two spherical
massive shells in a generally spherically symmetric background,
obtaining an expression from the conservation law that imposes a
constraint between the different parameters involved. We study the
light-like limit and make some comparisons of the predictions of our
master equation with the results obtained in the case of collision of
light-like shells, like the short life of white holes or the mass
inflation phenomena. We present some particular cases of the
constraint equation.}
\end{abstract}
\vspace{7mm}
\vfill
\eject

\section{Introduction}

The dynamics of a spherical shell expanding or contracting subject to
a gravitational pull in an given background space is an interesting
problem with applications in astrophysics, for instance the ejection
of matter in the explosion of a supernova\cite{Sato}, and in
cosmology to model the structure formation of galaxies and
clusters\cite{Benzin}. The dynamics of a shell moving in the
background of a charged black hole were beautifully discussed in the
pioneering work of Israel\cite{Israel1} and de la Cruz and
Israel\cite{Israel}. These studies showed that it is in principle
possible for a body collapsing to a black hole to re-emerge from a
white hole in a different universe, so  the analytical extensions of
a charged black hole to a white one appeared as a result of the
different phenomena view by the different observers, for the observer
at infinity the shell just go to the hole and is infinitely
red-shifted as it asymptotically approaches the exterior horizon, on
the other hand, for the observer falling with the shell, it takes a
finite amount of time to reach not only the horizon but further and
then re-bounce under some circumstances.

The case of two shells colliding had even more surprises. It is
remarkable that the predictions of new phenomena emerge from the good
old energy-momenta conservation law; everybody knows that energy and
momentum are the same before and after the collision, but an
important thing to notice is that for the contracting shell there is
a potential energy with respect to the expanding shell and after
collision this potential energy is all of a sudden removed, so it has
to transform into kinetic energy, generating therefore the known blue
shift effect. For the expanding shell the changes are as radical;
from an expansion in a space determined only by the gravitational
mass located at the origin, say $m_D$, after collision an observer
moving with the shell finds himself bound to a greater gravitational
mass: $m_D$ plus the gravitational mass of the imploding shell which
has increased. The changes can be so dramatic that the observer may
find himself inside a newly generated black hole!.

Some of the implications of these phenomena have been the conclusion
that the white holes have a short life\cite{Eardley,Blau}, (they are
buried by a black hole), and the mass inflation
phenomenon\cite{P&I:90,Israel2}, which is opening a new field of
research on the black holes interiors, for a review about these
see\cite{Israel3}.

The studies mentioned above have been done using null shells as a
model, there remained the question whether or not the conclusions
reached were still  valid in a more realistic model using one or both
shells massive. This is the problem with which we are concerned in
this paper. In the next section we will present a brief review of the
dynamics of massive spherical thin shells moving in curved
backgrounds. In the third section we use the energy-momentum
conservation law and arrive at a general constraint equation which
relates the different parameters involved in the collision. We prove
that when going to the light-like case we recover the known results,
also we present the case of  collision between a light-like shell and
a massive one, we also work with some particular cases for the
collision of two massive shells. Finally we present our conclusions
and suggest some lines of further research.

\section{Massive shells}
The junction conditions for arbitrary boundary surfaces and the
equations of motion for a time-like thin shell have been well
understood since the works of Israel\cite{Israel1},
Barrab\`es\cite{Barrabes}, de la Cruz and Israel\cite{Israel},
Kuchar\cite{Kuchar}, Chase\cite{Chase}, Balbinot and
Poisson\cite{BB}, and Lake\cite{Lake}. In this section we give a
brief review of this subject and obtain the equations of motion for a
thin shell.

Let $\Sigma$ be a hypersurface separating a given space-time M into
two parts, $M^\pm$, and ${n^\alpha}$ be the unit normal vector of
$\Sigma$ pointing from ${M^-}$ to ${M^+}$. Let ${x^\alpha_{\pm}}$ be
a system of coordinates in ${M^{\pm}}$ and let ${\xi^a}$ be a system
of intrinsic coordinates on $\Sigma$. The vectors ${e^\alpha_{(a)}}$
tangent to $\Sigma$ are defined by

\begin{equation}
e^\alpha_{(a)}=\frac{\partial x^\alpha}{\partial \xi^a},
\end{equation}
\begin{equation}
n_\alpha \, e^\alpha_{(a)}=0,
\end{equation}
and act as projectors from M onto the hypersurface $\Sigma$; from now
on we will suppress, in general, the use of the $ \pm$ indices.

It can be shown that the covariant derivative of any vector
$A^\alpha$ tangent to $\Sigma$ have components along the vectors
${e^\alpha_{(a)}}$, ${n^\alpha}$ and is given by:
\begin{equation}
A^\alpha_{|\beta}
\,e^\beta_{(b)}=A^a_{;b}e^\alpha_{(a)}+A^a\,K_{ab}\,n^\alpha,
\label{eq:A}

\end{equation}
where $A^\alpha=A^a \, e^\alpha_{(a)}$, the stroke denotes covariant
derivative with respect to the four-metric ${g_{\mu\nu}}$, and the
semicolon denotes covariant derivative with respect to three-metric
${h_{ab}=e^\alpha_{(a)}e_{\alpha(b)}}$. The extrinsic curvature
${K_{ab}}$ is defined by
\begin{equation}
K_{ab}=n_{\alpha|\beta} \, e^\alpha_{(a)}e^\beta_{(b)}=-n_\alpha \,
e^\alpha_{(a)|\beta} \, e^\beta_{(b)}.\label{eq:k}
\end{equation}

Taking $A^\alpha $ as the basis vector $e^\alpha_{(d)}$ in
(\ref{eq:A}) and calculating the covariant derivative along the
vector $e^\beta_{(c)}$ and using the  Ricci commutation relations we
can obtain the well-known Gauss-Codazzi equations
\begin{equation}
R_{\alpha \beta \gamma \delta} \, e^\alpha_{(a)} \,e^\beta_{(b)}
\,e^\gamma_{(c)} \,e^\delta_{(d)} = R_{abcd}- K_{ac} \, K_{bd} +
K_{bc}  \, K_{ad}.\label{eq:B}
\end{equation}

\begin{equation}

R_{\alpha \beta \gamma \delta} \,n^\alpha \,e^\beta_{(b)}
\,e^\gamma_{(c)} \,e^\delta_{(d)} =K_{bc;d} - K_{bd;c}.\label{eq:C}
\end{equation}

Acting on (\ref{eq:B}) and (\ref{eq:C}) with $ h^{a b}$ and using the
relation
\begin{equation}
h^{ab} \, e^\alpha_{(a)} \,e^\beta_{(b)} = g^{\alpha \beta} -n^\alpha
\, n^{\beta},
\end{equation}
we find

\begin{equation}
{}^3R - K_{ab} \,K^{ab} + K^2 =-2 G_{\alpha \beta} \,n^\alpha
\,n^\beta ,
\end{equation}
\begin{equation}
K^b_{a;b} - K_{;a} = -G_{\alpha \beta}\, e^\alpha_{(a)} \,n^\beta ,
\end{equation}
where ${}^3R $ is the intrinsic 3-curvature invariant of $\Sigma $, $
K=h^{ab} \,K_{ab}$ and $ G_{\alpha \beta }$ is the Einstein tensor.

In general, the components of the tensor $K_{ab} $ when measured with
respect to $M^+$ and $ M^- $ may be different. In order that $\Sigma
$ be the history of a thin shell we must impose the condition that
\begin{equation}
\gamma_{ab}=K^+_{ab} - K^-_{ab} = [K_{ab}],\label{eq:gam}
\end{equation}
 be non-vanishing. Such discontinuity is related to the intrinsic
stress energy

tensor of the surface, $S_{ab}$, through the Lanczos
equation\cite{Israel1}:
\begin{equation}
\gamma_{ab} - \gamma \, g_{ab} = - 8\,\pi\,S_{ab},\label{eq:S}
\end{equation}
where $g_{ab}$ is the intrinsic metric of $\Sigma$ and $\gamma =
g^{ab}\,\gamma_{ab}$.

The proper surface density, $\sigma$, is defined by the following
equation:
\begin{equation}
\sigma =  {S_a}^b\,u^a\,u_b ,\label{eq:sig}
\end{equation}
where $u^a$ is the vector tangent to the shell, already normalized,
$u^a\,u_a = -1$. From (\ref{eq:S}) and (\ref{eq:sig}), we can easily
find that
\begin{equation}
\gamma_{ab}\,u^a\,u^b + \gamma  =  - 8\,\pi\,\sigma.\label{eq:Ssig}
\end{equation}

Now we will consider the case of a spherical shell where the
intrinsic coordinates on $\Sigma$ are given by $\xi^a=(\tau, \theta,
\phi)$; $\tau$ is the proper time along the streamlines
$\theta,\phi=const.$ so the line element

$ds^2|_\Sigma$ is given by
\begin{equation}
ds^2|_\Sigma=R^2(\tau) \, d\Omega^2 - d\tau^2, \label{eq:Slin}
\end{equation}
where $d\Omega^2=d\theta^2 + \sin^2\theta \,d\phi^2$, and $R(\tau )$
is the radius of the shell. According to the Birkhoff's theorem, the
line element in both $M^+$ and $M^-$ is reducible to
\begin{equation}
ds^2_\pm= {\cal H}_\pm\,dv_\pm \, ( - {\cal H}_\pm\,f_\pm\,dv_\pm  +
2\, dr) + r^2 \, d\Omega^2, \label{eq:ell} \end{equation}
where ${\cal H}_\pm$ and $f_\pm$ are functions of $r$, and the
respective $v_\pm$.

The unit normal and velocity are easily calculated and have the
following expression:
\begin{equation}
n_\alpha = \epsilon \, {\cal H}\,(-\dot R, \dot v, 0, 0),
\label{eq:nor}
\end{equation}
\begin{equation}
u^\alpha = (\dot v, \dot R, 0, 0). \label{eq:vel},
\end{equation}
where we have introduced the factor $\epsilon=\pm 1$ to indicate the
increasing or decreasing of the radius $r$ along the normal.

At this point we are able to deduce the motion equation for $\Sigma$
in a direct and general way. Following Lake\cite{Lake}, from
eqs.~(\ref{eq:Ssig}) and (\ref{eq:Slin}) it follows that
\begin{equation}
\gamma_{\theta \theta}= - 4\,\pi R^2\,\sigma,
\end{equation}
notice that the right hand side is the proper mass of the shell,
$M(\tau )$
\begin{equation}
4\,\pi R^2\,\sigma = M(\tau ).
\end{equation}

Now, with this last equation and recalling the definition of the
$\gamma_{\theta \theta}$, eq.~(\ref{eq:gam}), we obtain that
\begin{equation}
{K_{\theta \theta}}^+ - {K_{\theta \theta}}^- = M(\tau
).\label{eq:k1}
\end{equation}

It proves convenient to rewrite (\ref{eq:k1}) as follows
\begin{equation}
{K^2_{\theta \theta}}^+ = \frac1{4 M^2(\tau )}\left({K^2_{\theta
\theta}}^- -
{K^2_{\theta \theta}}^+  - M^2(\tau ) \right)^2.\label{eq:kfin}
\end{equation}

{}From the line element of a general spherically symmetric space,
eq.~(\ref{eq:ell}), we can calculate the extrinsic curvature using
eq.~(\ref{eq:k}), to obtain
\begin{equation}
K_{\theta \theta}= \epsilon R(\tau ) \left( {\cal H}\,f\, \dot v -
\dot R \right),
\end{equation}
also, equating the line element eq.~(\ref{eq:ell}) at both sides of
the surface with the line element at the surface, $ds^2_\pm|_\Sigma =
ds^2|_\Sigma$, we obtain that
\begin{equation}
{\cal H}_\pm\,\dot v_\pm ({\cal H}_\pm\,f_\pm\,\dot v_\pm - 2 \dot R
)= 1.
\end{equation}

Substituting these last two equations in eq.~(\ref{eq:kfin}) we
obtain the following motion equation:
\begin{equation}
{\dot R}^2 = (\frac{R}{2\,M})^2\,(f_+ - f_-)^2 - \frac12 (f_+ + f_-)
+
(\frac{M}{2\,R})^2. \label{eq:mot}
\end{equation}

We want to make some remarks about this motion equation: first as
long as in the deduction of this motion equation the explicit form of
the stress energy tensor of the shell, $S_{ab}$, was not used,  then
this result is valid for any spherically symmetric thin shell;
second, it is an equation valid for any spherically symmetric
background, eq.~(\ref{eq:ell}). Finally, we also want to call the
attention to the fact that it was not necessary to obtain the
equation for the acceleration, $\ddot R(\tau )$, and then make a
first integration, as it is done in other derivations of the motion
equation of a shell.

Before going into the implications of this relation in the case of
collision, we think that going to the Schwarzschild case can clarify
the meaning of the terms that appear in the equation(\ref{eq:mot}).
In this case $f_\pm=1 - \frac{2m_{1/2}}R$, so eq.~(\ref{eq:mot}) can
be put in the following form:
\begin{equation}
m_2 - m_1 =M(1-\frac{2m_1}{R}+{\dot R}^2)^{1/2}- \frac{M^2}{2 \,
R^2}.
\end{equation}
This last equation expresses the total gravitational mass of the
shell; expanding the squared root to first order it can be reduced to
a sum of four well known terms: the proper mass of the shell, $M$,
the kinetic energy $\frac M2{\dot R}^2 $, the mutual potential energy

$ -\frac{M \, m_1}R $ and a self-potential energy $-\frac{M^2}{2\, R}
$.

In our specific case we are dealing with two concentric spherical
thin massive shells colliding without interaction and propagating in
the field due to a spherically symmetric mass distribution $m_D$ near
its centre (fig. 1). In this case the space-time is separated into
four radial sectors A, B, C, D and the gravitational mass of the
in-falling shell is given by  the  difference $m_C-m_B$ and  that of
the outgoing shell by $m_B - m_D$. The equations of motion for the
two shells before collision are
\begin{eqnarray}
\dot {r_{IV}}^2 & = & (\frac{r_{IV}}{2M_{IV}})^2\,(f_B - f_D)^2
-\frac12 (f_B + f_D) + (\frac{{M_{IV}}}{2\,{r_{IV}}})^2,\nonumber \\
\dot r_{III}^2 & = & (\frac{r_{III}}{2M_{III}})^2\,(f_C - f_B)^2
-\frac12 (f_B + f_C) + (\frac{{M_{III}}}{2\,{r_{III}}})^2, \nonumber
\\\label{eq:mota}
\end{eqnarray}
and after collision
\begin{eqnarray}
\dot {r_{II}}^2 & = & (\frac{r_{II}}{2M_{II}})^2\,(f_C - f_A)^2
-\frac12 (f_A + f_C) + (\frac{{M_{II}}}{2\,{r_{II}}})^2, \nonumber \\
\dot {r_I}^2 & = &(\frac{r_{I}}{2M_{I}})^2\,(f_A - f_D)^2 -\frac12
(f_A + f_D) + (\frac{{M_{I}}}{2\,{r_{I}}})^2,\nonumber
\\\label{eq:motb}
\end{eqnarray}
where $M_i (i=I,...,IV)$ are the rest mass of the shells.

\section{conservation relation for the collision of massive shells}
\setcounter{equation}{0}

The energy-momentum conservation law can be used to obtain a relation
the energies and momenta of the different regions before and after
the collision. This problem has been studied by several authors as
Redmount\cite{Redmount}, Dray and 't Hooft\cite{DH} and Barrab\`es,
Israel and Poisson\cite{P&I:90,BIP}, among others and they have
obtained an expression for the case of collision between null shells
and has proved to be central in further studies as we mentioned in
the introduction. Now we proceed to derive a generalization of such a
relation in the case of collision of massive spherical shells.

We have four surfaces, two after and two before the collision. The
normal vector to the respective surface is given by eq.
(\ref{eq:nor}), which can be expressed as
\begin{equation}
n_\alpha = \epsilon {\cal H}(- \dot R ,\frac{\dot R \pm \sqrt{f +
{\dot R}^2}}{f\,{\cal H}} , 0,0), \label{eq:n}
\end{equation}
where we have use the line element equation (\ref{eq:ell}), evaluated
at the surface to obtain a relation between the velocities $\dot t$
and $\dot R$.
In order to proceed, we need the velocities of the shells,
$u^\alpha$, given by eq. (\ref{eq:vel}),  but we remark that these
velocities can be normalized with respect to one or the other
adjacent region, so we finally have 8 velocities, ${u^\alpha}_i
|_{\pm} = \frac{d{x^\alpha}_i}{d\tau_i} |_{\pm}$, with plus (minus)
stands for the space to the right (left) of the shell {\it in the
direction of the motion}, see figure 1!:
\begin{eqnarray}
{u^\alpha}_I |_\pm & = & (\frac{\dot {r_{I}} - \sqrt{f_{A/D} + {\dot
r_{I}}^2}}{f_{A/D}\,{\cal H}_{A/D}}  , \dot r_I , 0, 0), \nonumber \\

{u^\alpha}_{II} |_\pm & = & (\frac{\dot {r_{II}} + \sqrt{f_{C/A} +
{\dot r_{II}}^2}}{f_{C/A}\,{\cal H}_{C/A}} , \dot r_{II} , 0, 0),
\nonumber \\
{u^\alpha}_{III} |_\pm & = & (\frac{\dot {r_{III}} + \sqrt{f_{C/B} +
{\dot r_{III}}^2}}{f_{C/B}\,{\cal H}_{C/B}}, \dot r_{III} , 0, 0),
\nonumber \\
{u^\alpha}_{IV} |_\pm & = & (\frac{\dot {r_{IV}} - \sqrt{f_{B/D} +
{\dot r_{IV}}^2}}{f_{B/D}\,{\cal H}_{B/D}} , \dot r_{IV} , 0, 0),
\nonumber \\
\end{eqnarray}
$f_{M/N}$ corresponds to the respective $\pm$.

The relations between the spacetime angles with which the shells
collided and went on, can be given in terms of a relation for the
scalar products between the different normals, and we have four
different such products:
\begin{eqnarray}
u_{IV}|_+ \cdot u_{III}|_- & =  {g_{\alpha \beta}}_B {u_{IV}}^\alpha
|_+ {u_{III}}^\beta |_-   & =

 \frac{\sqrt{( f_B + {\dot r_{IV}}^2)( f_B + {\dot r_{III}}^2)} +
\dot r_{IV}\dot r_{III}}{f_B} ,\nonumber \\
u_{I}|_+ \cdot u_{II}|_- & =  {g_{\alpha \beta}}_A {u_{I}}^\alpha |_+
{u_{II}}^\beta |_- & = \frac{\sqrt{( f_A + {\dot r_{I}}^2)( f_A +
{\dot r_{II}}^2)} +
\dot r_{I}\dot r_{II}}{f_A} , \nonumber \\
 \label{dotp}
\end{eqnarray}
analogously, the relation between the angles can be given in terms of
the scalar products in the other two regions, C and D. Notice how in
eq.~(\ref{dotp}) the metric coefficient ${\cal H}$ of the different
regions does not appear at all.

Now, from the conservation of the 4-momentum

\begin{equation}
{p^\mu}_I + {p^\mu}_{II} = {p^\mu}_{III} + {p^\mu}_{IV},
\end{equation}
and considering a collision in which the interaction is purely
gravitational, so the rest mass of the shell after the collision is
the same as before
\begin{equation}
M_I=M_{III}, \mbox{\hspace{.25in}} M_{II}=M_{IV},\label{eq:mas}
\end{equation}
the conservation of momenta implies the following relation between
the velocities

\begin{equation}
u_{I}|_+ \cdot u_{II}|_- = u_{IV}|_- \cdot u_{III}|_- ,
\end{equation}
and using (\ref{dotp}) we obtain that

\begin{equation}
 \frac{\sqrt{(f_A + {\dot r_{I}}^2)( f_A + {\dot r_{II}}^2)} +
\dot r_{I}\dot r_{II}}{f_A}|_c  =

\frac{\sqrt{(f_B + {\dot r_{IV}}^2)( f_B + {\dot r_{III}}^2)} +

\dot r_{IV}\dot r_{III}}{f_B}|_c  \label{eq:non}
\end{equation}
we remark that this relation is evaluated at the collision 2-surface.

In order to proceed further, we have to take in account the motion
equations, eq.~(\ref{eq:mota}) and eq.~(\ref{eq:motb}) for each
region, which it proves helpful to rewrite as:
\begin{eqnarray}
{\dot r_{I}}^2 & = & \frac{{{\cal R}_I}^2}
{4 {r_I}^2 {M_{I}}^2} - f_A, \nonumber \\
{\dot r_{II}}^2 & = &  \frac{{{\cal R}_{II}}^2}
{4 {r_{II}}^2 {M_{II}}^2} - f_A, \nonumber \\
{\dot r_{III}}^2 & = &  \frac{{{\cal R}_{III}}^2}
{4 {r_{III}}^2 {M_{III}}^2} - f_B, \nonumber \\
{\dot r_{IV}}^2 & = & \frac{{{\cal R}_{IV}}^2}
{4 {r_{IV}}^2 {M_{IV}}^2} - f_B,
\end{eqnarray}
where we have defined
\begin{eqnarray}
{\cal R}_{I}& = &  {r_I}^2(f_D - f_A ) - {M_{I}}^2, \nonumber \\
{\cal R}_{II} & = &  {r_{II}}^2(f_C - f_A ) - {M_{II}}^2, \nonumber
\\
{\cal R}_{III} & = &  {r_{III}}^2 (f_C - f_B ) - {M_{I}}^2, \nonumber
\\
{\cal R}_{IV} & = & {r_{IV}}^2(f_D - f_B ) - {M_{II}}^2.

\label{eq:ri}
\end{eqnarray}
Now we can work with the conservation relation (\ref{eq:non}),
remembering that all the $r_i$ are $r_c$, the radius of collision,
and that the respective masses are equal, eq.~(\ref{eq:mas}), after
some manipulation we can rewrite (\ref{eq:non}) as :
\begin{equation}
 {\cal A}[{\cal R}_{III}\, {\cal R}_{IV} (\alpha   + \beta  ) -
{\cal R}_{I}\, {\cal R}_{II}( \gamma + \beta )]  =

{{r_c}}^2\,(f_B \alpha - f_A \gamma + (f_A - f_B)\beta )^2   ,
\label{eq:cc}
\end{equation}
where
\begin{eqnarray}
\alpha & = & {r_c}^4\,[{M_{I}}^2(f_C - f_A)^2 + {M_{II}}^2(f_D -
f_A)^2], \nonumber \\
\gamma & = & {r_c}^4\,[{M_{I}}^2(f_D - f_B)^2 + {M_{II}}^2(f_C -
f_B)^2], \nonumber \\
\beta & = & {M_I}^2 {M_{II}}^2[{M_I}^2 + {M_{II}}^2 - 2 {r_c}^2 (f_C
+ f_D) ],
\nonumber \\
{\cal A} & = & {\cal R}_{I}\, {\cal R}_{II} f_B - {\cal R}_{III}\,

{\cal R}_{IV} f_A.
\end{eqnarray}
and with ${\cal R}_i$ defined in (\ref{eq:ri}), evaluated at $r_c$.

Equation (\ref{eq:cc}) is the constraint equation for the collision
of two massive spherical shells with arbitrary stress energy tensor,
in a general spherically symmetric background, under the assumption
that the collision is transparent, so the interaction is purely
gravitational.

In order to go to the light-like limit, we proceed as follows: let us
make one of the proper masses, say $M_I$ equal to zero, so after some
simplifications we obtain
\begin{eqnarray}
&(f_C - f_B)\,(f_D - f_A)[(f_C + f_D) (f_A + f_B - f_C -
f_D)\,{r_c}^2 + (f_A - f_B + f_C - f_D)\,{M_{II}}^2]\times &
\nonumber \\

&[(f_A - f_B)\,(f_A\,f_B - f_C\,f_D)\,{r_c}^2  + (f_A\,f_C - f_B\,f_D
)\,{M_{II}}^2] =& \nonumber \\

&[(r_c\,M_{II})^2\,[f_B\,(f_D - f_A)^2 -f_A\,(f_C - f_B)^2]^2, &
\label{eq:mix}\end{eqnarray}
which represents the conservation equation in the case of collision
between a null shell, in this case the imploding one, with a massive
exploding one.  Below we  analyse more about this case, to proceed
further with the light-light limit now we set the other proper mass,
$M_{II}$, equal to zero so we obtain that

\begin{equation}
(f_C - f_B)\,(f_D - f_A)\,(f_C + f_D)\,(f_A + f_B - f_C - f_D)(f_A -
f_B)(f_A\,f_B - f_C\,f_D) = 0,
\end{equation}
which is the conservation law for the case of collision of two
massiveless shells. Now, if we suppose that there are actually
shells, then always

$f_C \neq f_B$, $f_A \neq f_D$, and always $f_C + f_D \neq 0$, so the
last equation implies that
\begin{equation}
(f_A - f_B)(f_A + f_B - f_C - f_D)(f_A\,f_B - f_C\,f_D) = 0,
\end{equation}
which implies that either one, two or the three factor are equal to
zero, that is
\begin{equation}
f_A\,f_B = f_C\,f_D , \label{eq:null}
\end{equation}
or
\begin{equation}
f_A= f_B, \label{eq:nulla}
\end{equation}
or
\begin{equation}
f_A + f_B = f_C + f_D,\label{eq:nullb}
\end{equation}
equation (\ref{eq:null}) is the known result for the collision of non
massive shells\cite{P&I:90,Redmount,DH}. What comes as a surprise is
the fact that the conservation relation in the light-like limit also
would be satisfied if eq.~(\ref{eq:nulla}) or eq.~(\ref{eq:nullb})
holds. This fact implies that under the circumstances given by these
two equations, the known result (\ref{eq:null}) does not necessarily
have to be satisfied.

As long as many works on the light-light shells collision have been
done in the Schwarzschild background, we found interesting to study
in more detail this particular case of our master equation
(\ref{eq:cc}). Using the fact that for Schwarzschild $f = 1 - \frac{2
m}r$ and ${\cal H} = 1$ in eq.~(\ref{eq:ell}), it can be proved that
in this case equation (\ref{eq:cc}) is in general a sixth order
polynomial in the radius of collision, $r_c$, (actually, seventh
order but one of the roots is  $r_c=0$), where the coefficients are
functions of the rest of the parameters, namely the four
gravitational masses and the two proper masses. However, in the case
of a collision between a light-like shell and a massive one, either
the null shell expanding and the massive contracting or viceversa,
equation (\ref{eq:cc}) reduces to a third order polynomial in $r_c$.
The case when the proper masses are the same, $M_I=M_{II}$, also
produces some simplification in the equation (\ref{eq:cc}), reducing
the order to fourth. We recall that the light-light collision reduces
to a first order polynomial in $r_c$, namely
\begin{equation}
r_c=\frac{2(m_A\,m_B - m_C\,m_D)}{m_A + m_B - m_C -m_D}.

\end{equation}

Using  numerical analysis we studied the dependence of the radius of
collision with respect to the proper masses, for fixed values of the
gravitational ones, that is, the coefficients in eq.(\ref{eq:cc})
were only functions of $M_I$ and $M_{II}$, so we solved it
numerically. We worked with different values of the gravitational
masses and the results seem to point to the fact that the radius of
collision for the light-light shells collision is a maximum; we
present an example showing this result in figure 2, the chosen values
of the gravitational masses in this figure were, $m_A = 3$, $m_B =
5$, $m_C = 7$, $m_D = 0$, the radius of collision for the light-light
case is $r_c = 30$, notice how this is a maximum.

Now, returning to the master equation (\ref{eq:cc}), let us discuss
something about the mass inflation, for details about it
see\cite{Israel2,P&I:90}. According to this model, inside the charged
black hole there is the cross flux of two null shells, the incoming
one is parallel and very close to the inner horizon, which in the
unperturbed case is also the locus of the Cauchy horizon. In figure
1, line I-III would represent this in-flux. In this way, the
collision is supposed to happen  close to the inner horizon of a
charged black hole, so in (\ref{eq:null}) $f_C$ is going to zero, but
$f_A$ and $f_B$ remain finite, so does their product, which due to
the equality, implies that the product $f_C\,f_D$ has to remain
finite, which implies in turn that $f_D$ has to diverge. This is a
practical way to see the mass inflation.

Now, as written, our master equation, eq.~(\ref{eq:cc}), does not
have the kind of behavior mentioned for the null-null case; it is not
expressed as an equality between products that would repeat the
procedure mentioned above and so there is no way to conclude that
some factors or coefficients should diverge in order to maintain the
equality.

Also, during the collision between a light-like shell and a massive
one, eq.~(\ref{eq:mix}), although it is simpler, as it is, either has
the mentioned behavior of the light light case, so in order to say
something about the mass inflation for the massive case, it is better
to do a thin-shell analysis as done by Brady et. al\cite{Brady}, this
analysis seems to point at the fact that for a null shell incoming
parallel to the inner horizon and colliding with a massive one in the
interior of a charged black hole, there would be a mass inflation
described in the leading terms by light-light case; the respective
massive-massive collision is still under study.

\section{Conclusions}
The main conclusion of the present work is that the results obtained
from the analysis of the collision of null shells are, in general,
valid, collecting the leading behavior of the master equation
(\ref{eq:cc}). With respect to white holes, it seems that the
collision point for null shells is a maximum, so the conclusions
reached by Eardley\cite{Eardley} do describe the leading phenomena in
the more complete analysis using the master equation, mainly that if
the white holes appear in a medium with matter, they will be buried
down very quickly by the black hole formed after the collision. The
same seems to apply for the mass inflation phenomena: the leading
behavior is correctly described by the null case.

Nevertheless, we want to stress the fact that it is needed a more
realistic analysis of the collision, such as the one described in
this paper, to give a more solid ground to the former conclusion.

The fact of having a full master equation (\ref{eq:cc}) is important
to analyse particular cases which might provide us with some new
situations, for instance, the cases in which the known relation for
the null case is not necessarily satisfied, eqs.~(\ref{eq:nulla},
\ref{eq:nullb}), deserve a deeper study which is currently on its
way.

Finally, we want to remark that the master equation obtained,
eq.~(\ref{eq:cc}), is valid for the most general spherically
symmetric background and that it applies for a wide type of massive
thin shells, as long as the specific form of the stress energy tensor
of the shell is not used in the derivation of the master equation
(\ref{eq:cc}).

\section{Acknowledgements}
It is a great pleasure to thank Werner Israel for suggesting the
problem  and for fruitful discussions and encouragement, we are also
grateful to Patrick Brady for comments and suggestions. D. N. is
grateful for the awards given by External Affairs and International
Trade Canada, Government of Canada Awards, administrated by the
International Council for Canadian Studies.  He also thanks DGAPA and
UNAM for support.  This work was also partly supported by the Natural
Sciences and Engineering Research Council of Canada. H. de O. would
like to acknowledge CAPES for financial support,  and J. S. does it
to CNPq. Figure~2, and some calculations in this work were done using
Mathematica$^{\copyright}$, we are grateful to A. Macpherson and S.
Droz for useful suggestions during the calculations.

\figure{Two views of the colliding shells: before and after the
collision.}

\figure{Point of collision as a function of the proper masses, for
the gravitational masses of $m_A=3$, $m_B=5$, $m_C=7$, $m_D=0$. For
these values, the point of collision in the light-light case is 30.}

\end{document}